\documentclass{article}
\usepackage{spconf,amsmath,graphicx}
\usepackage{bm, amssymb, booktabs, multirow,cite, url, color}

\title{MULTI-CHANNEL TARGET SPEECH EXTRACTION WITH CHANNEL DECORRELATION AND TARGET SPEAKER ADAPTATION}

\name{Jiangyu Han$^{1}$, Xinyuan Zhou$^{1}$, Yanhua Long$^{1}$\sthanks{Yanhua Long is the corresponding author. The work is supported by the National Natural Science Foundation of China (Grant No.62071302 and No.61701306).}, Yijie Li$^2$}
\address{
  $^1$Shanghai Normal University, Shanghai, China\\
  $^2$Unisound AI Technology Co., Ltd., Beijing, China}

\begin{document}
\ninept
\maketitle
\begin{abstract}

The end-to-end approaches for single-channel target speech extraction
have attracted widespread attention. However, the studies for end-to-end
multi-channel target speech extraction are still relatively limited.
In this work, we propose two methods for exploiting the multi-channel spatial information to extract the target speech.
The first one is using
a target speech adaptation layer in a parallel encoder architecture.
The second one is designing a channel decorrelation mechanism
to extract the inter-channel differential information to enhance
the multi-channel encoder representation. We compare the proposed methods with
two strong state-of-the-art baselines. Experimental results on the
multi-channel reverberant WSJ0 2-mix dataset demonstrate that
our proposed methods achieve up to $11.2\%$ and $11.5\%$ relative improvements in SDR and SiSDR respectively, which are the best reported results on this task to the best of our knowledge.

\end{abstract}


\section{Introduction}
\label{sec:intro}

Speech separation (SS) is a task that aims to separate each source signal
from the mixed speech. Many effective methods have been proposed to perform
the separation in the time-frequency domain, such as
deep clustering \cite{dpcl}, deep attractor network \cite{dan},
and permutation invariant training (PIT) \cite{pit}, etc.
More recently, a convolutional time-domain audio separation network
(Conv-TasNet)\cite{tasnet} has been proposed and achieved significant separation
performance improvement over those time-frequency based techniques.
This Conv-TasNet has attracted widespread attention and further improved
in many recent works, either for the single-channel or
multi-channel speech separation tasks \cite{tasnet_relate1, tasnet_relate2, tasnet_relate3}.

Our work is also to improve the Conv-TasNet, however, instead of for the pure speech separation, we aim to generalize such an
idea to the target speech extraction (TSE) task.
Generally, compared with the pure speech separation techniques,
most TSE approaches require the additional target speaker clues to %guide the speech separation network to extract only the speech of target speaker.
drive the network towards extracting the target speech.

Many previous works have focused on the TSE task,
such as Voicefilter \cite{voicefilter}, SBF-MTSAL-Concat \cite{sbf}, and SpEx \cite{spex}.
Although these close-talk TSE approaches achieve great progress, the performance of far-field speech extraction is still far from satisfactory due to the reverberation.
When the microphone arrays are available, the additional multi-channel spatial information usually can be helpful in the
TSE task. Such benefits have attracted many studies to
exploit the multi-channel information.
For example, the direction-aware SpeakerBeam \cite{direct}
combines an attention mechanism with beamforming to enhance
the signal of the target direction;
the neural spatial filter \cite{neural} use the directional information of the target speaker to extract the corresponding speech; the time-domain SpeakerBeam (TD-SpeakerBeam) \cite{td_speakerbeam} incorporates the inter-microphone phase difference (IPD)\cite{ipd} as additional input features
to further improve the speaker discrimination capability.
All of these multi-channel TSE approaches showed promising results,
which indicates that the multi-channel information can provide an alternative guider
to discriminate the target speaker better.
Actually, one of the keys of the TSE is still speech separation.
In order to further enhance the separation ability,
many strategies for exploiting the multi-channel information have been recently proposed,
such as normalized cross-correlation (NCC) \cite{ncc}, transform-average-concatenate (TAC) \cite{tac}, and inter-channel convolution difference (ICD)\cite{icd}, etc.
Therefore, how to effectively exploit the multi-channel spatial information for TSE is crucial.

In this study, we also focus on exploiting the multi-channel spatial
information for the target speech extraction task. Two approaches are proposed:
first, we integrate a target speech adaptation layer
into a parallel encoder architecture. This adaptation is used to enhance the
target speaker clues of multi-channel encoder output by weighting
the mixture embeddings.
Second, unlike creating hand-crafted spatial features,
we design a channel decorrelation mechanism
to extract the inter-channel differential spatial information automatically.
This decorrelation is performed on each dimension of all the multi-channel
encoder representations of input mixtures. Furthermore, together with the
same target speech adaptation layer, the proposed decorrelation mechanism can
provide another effective representation of the target speaker to enhance the whole end-to-end TSE network. To validate the effectiveness of our proposed approaches,
we choose two state-of-the-art TSE systems as the baselines, one is the
parallel encoder proposed in \cite{para} and the other is the
TD-SpeakerBeam with IPD features in \cite{td_speakerbeam}.
All of our experiments are performed on the publicly
available multi-channel reverberant WSJ0 2-mix dataset.
Results show that our proposed methods improve
the performances of multi-channel target speech
extraction significantly. Moreover, to make the research reproducible,
we released our source code on GitHub \footnote{https://github.com/jyhan03/channel-decorrelation}.

\section{TIME-DOMAIN SPEAKERBEAM}
\label{sec:td_speakerbeam}

\begin{figure}[t]
  \centering
  \includegraphics[width=8.0cm]{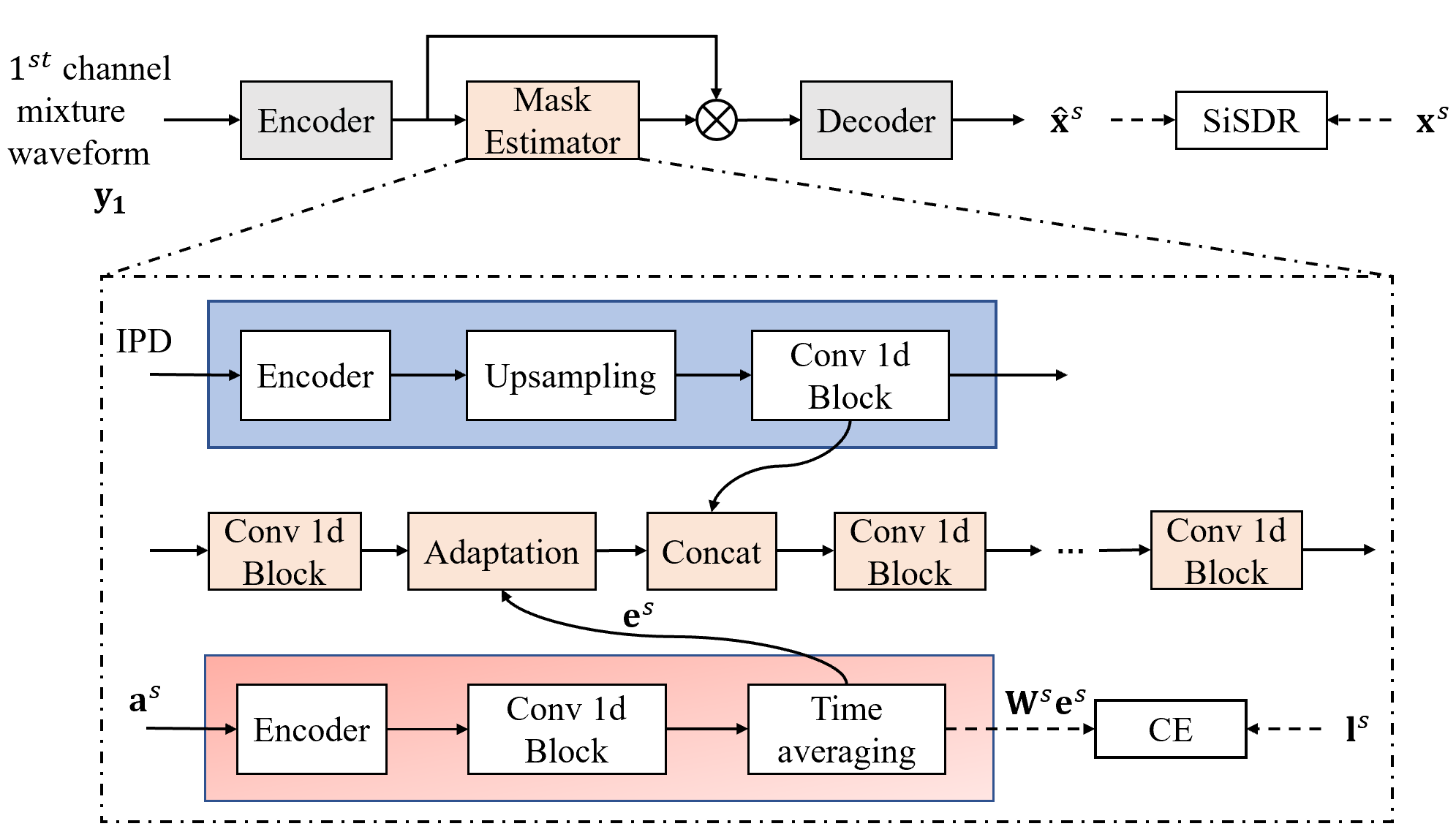}
  \caption{The block diagram of the TD-SpeakerBeam with IPD.}
  \label{fig:td_speakerbeam}
\end{figure}

TD-SpeakerBeam is a very effective
target speech extraction approach that has been recently
proposed in \cite{td_speakerbeam}.
The structure of TD-SpeakerBeam is shown
in Fig.\ref{fig:td_speakerbeam} without the IPD concatenation block.
It contains three main parts:
encoder (1d convolution layer), mask estimator (several convolution blocks),
and decoder (1d deconvolution layer). The $\mathbf{y_1}$,
$\mathbf{\hat{x}}^s$, and $\mathbf{a}^s$
are the mixture waveform of the first (reference)
channel, the extracted target speech waveform, and the adaptation
utterance of the target speaker respectively. The TD-SpeakerBeam
network follows a similar configuration as Conv-TasNet\cite{tasnet},
except for inserting a multiplicative adaptation layer \cite{adap}
between the first and second convolution blocks to drive the network towards extracting the target speech.
The adaptation layer accepts the mixture embedding matrix and the
target speaker embedding vector $\mathbf{e}^s$ as inputs.
In the adaptation layer, $\mathbf{e}^s$ is repeated to perform element-wise multiplication with the mixture embedding.
The $\mathbf{e}^s$ is computed by a time-domain convolutional
auxiliary network as shown in the bottom of Fig.\ref{fig:td_speakerbeam}.

In TD-SpeakerBeam, the network accepts the time-domain
signal of the mixture and outputs the time-domain
signal of the target speaker directly.
Moreover, as shown in Fig.\ref{fig:td_speakerbeam}, authors in \cite{td_speakerbeam}
also extended the TD-SpeakerBeam to the multi-channel TSE task, 
however, they only simply concatenated the hand-crafted 
IPD features (processed with a 1d convolutional encoder, upsampling, and a convolution block) with the adapted encoder representation to exploit the multi-channel
spatial information.

The whole network architecture of TD-SpeakerBeam is trained jointly
in an end-to-end multi-task way. The multi-task loss combines
the scale-invariant signal-to-distortion ratio (SiSDR)\cite{sisdr} as the
signal reconstruction loss and cross-entropy as the
speaker identification loss. The overall loss function is defined as,
\begin{equation}\small
L(\Theta|\mathbf{y}, \mathbf{a}^s,\mathbf{x}^s,\mathbf{l}^s) = {\rm -SiSDR}(\mathbf{x}^s, \mathbf{\hat{x}}^s) + \alpha{\rm CE}(\mathbf{l}^s, \sigma(\mathbf{W^se}^s))
\label{eq:multi}
\end{equation}
where $\Theta$ represents the model parameters, $\mathbf{y}$ is the input mixture, $\mathbf{x}^s$ is the target speech signal, $\mathbf{l}^s$ is a one-hot vector representing the target speaker identity, $\alpha$ is a scaling parameter, $\mathbf{W^s}$ is a weight matrix, and $\sigma(\cdot)$ is a softmax operation.
More details can be found in \cite{td_speakerbeam}.

\section{PROPOSED METHODS}
\label{sec:cdb_all}

\subsection{Parallel encoder with speaker adaptation}
\label{ssec:parallel}

\begin{figure}[ht]
  \centering
  \includegraphics[width=9.0cm]{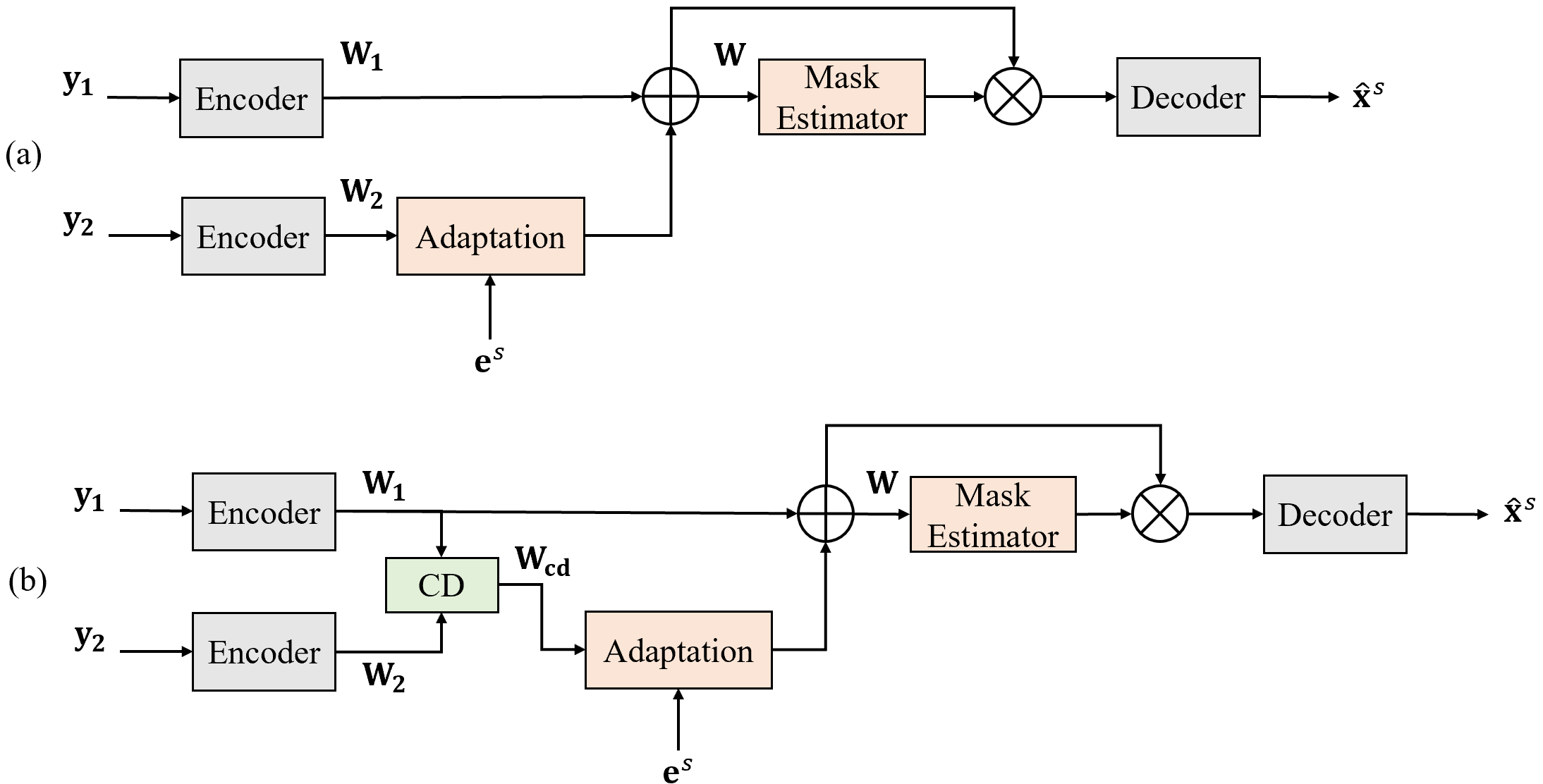}
  \caption{The proposed multi-channel TSE based on TD-SpeakerBeam with (a) parallel encoder with speaker adaptation; (b) channel decorrelation. $\mathbf{e^s}$ is the same as in Fig.\ref{fig:td_speakerbeam}. }
  \label{fig:cdb_related}
\end{figure}

Instead of using the IPD features, we first extend the
TD-SpeakerBeam to multi-channel and introduce a
target speaker adaptation layer
in a parallel encoder architecture that has been proposed
in \cite{para}. The original parallel encoder is shown
in Fig.\ref{fig:cdb_related} (a)
without the adaptation block. It just directly sums the
waveform encodings of each input channel to form the final
mixture representation. However, to enhance the target speaker
clues of multi-channel encoder output, besides inserting the adaptation layer
in the mask estimator of TD-SpeakerBeam,
we also integrate a same adaptation layer in the parallel encoder. The diagram is
shown in Fig.\ref{fig:cdb_related} (a).

\subsection{Channel decorrelation}
\label{ssec:cd}

Unlike exploiting the cross-correlations between multi-channel
speech in the original parallel encoder.
In this paper, we design a channel decorrelation (CD) mechanism to improve
the multi-channel TSE performance on TD-SpeakerBeam. The proposed
network structure is shown in Fig.\ref{fig:cdb_related} (b).
This decorrelation aims to extract the inter-channel differential
spatial information automatically. It is performed on each dimension
of all the multi-channel encoder representations of input
mixtures. Take the case of two channels as an example,
as shown in Fig.\ref{fig:cdb_related} (b), the
CD block accepts two encoded mixture representations $\mathbf{W_1}$
and $\mathbf{W_2}$, then outputs the differential spatial information
$\mathbf{W_{cd}}$ between two channels. The specific procedure is as follows:

First, compute the cosine correlation between $\mathbf{W_1}$
and $\mathbf{W_2}$ in each corresponding dimension (row).
Note that $\mathbf{W_1}$ and  $\mathbf{W_2}$ are two matrices, i.e.,
\begin{equation}
\mathbf{W_i} = {[\mathbf{w_i}^1, \mathbf{w_i}^2, ..., \mathbf{w_i}^N]}^\mathbf{T}, i = 1, 2
\end{equation}
where $\mathbf{W_i} \in \mathbb{R}^{N \times T}$ is the input
of the $i$-th channel of CD, $N$ is the output dimension
of convolutional encoder, $T$ is the number of frames of the encoder output.
$\mathbf{w_i}^j \in \mathbb{R}^{1 \times T}, j = 1, 2, ..., N$,
is the $j$-th dimension vector of the $i$-th channel.
$\mathbf{T}$ is the operation of transpose.

The cosine correlation of $j$-th dimensional vector between
the first and second channel is calculated as,
\begin{equation}
\phi_{1,2}^j = \frac{\langle \mathbf{w_1}^j, \mathbf{w_2}^j \rangle}{\lVert{\mathbf{w_1}^j}\rVert_2 \lVert{\mathbf{w_2}^j}\rVert_2},  j = 1, 2, ..., N
\end{equation}
where $\langle \cdot \rangle$ is the inner product of two vectors,
$\lVert\cdot\rVert_2$ represents the Euclidean norm.
The vectors involved in the operation are normalized to zero mean prior to the calculation.

Then, calculate the cosine correlation between the vectors of each
dimension of $\mathbf{W_1}$ and  $\mathbf{W_2}$ in turn, and concatenate them to
get a similarity vector $\bm{\phi_{1,2}}$,
\begin{equation}
\bm{\phi_{1,2}} = [\phi_{1,2}^1, \phi_{1,2}^2, ..., \phi_{1,2}^N]^\mathbf{T}
\end{equation}
where $\bm{\phi_{1,2}} \in \mathbb{R}^{N \times 1}$ represents the
similarity of two encoded mixture representations
in each dimension in a latent space.

Next, to convert each dimension of the similarity vector $\bm{\phi_{1,2}}$
into a probability vector $\mathbf{p_{1,2}}$, we introduce an
auxiliary vector $\mathbf{a}$ with the same size
as $\bm{\phi_{1,2}}$ and values are all 1, i.e.,
\begin{equation}
\mathbf{a} = [1,1,...,1]^\mathbf{T}
\end{equation}
where $\mathbf{a} \in \mathbb{R}^{N \times 1}$ can be regarded as the
cosine correlation between first channel $\mathbf{W_1}$ and itself or a probability vector whose values are all 1.
%which represents the reference signal of $\bm{\phi_{1,2}}$.
Then, a softmax operation is used to
calculate the probability between each corresponding
dimensional element in $\bm{\phi_{1,2}}$ and $\mathbf{a}$ to
obtain the final similarity probability vector $\mathbf{p_{1,2}} \in \mathbb{R}^{N \times 1}$, i.e., 
\begin{equation}
\begin{split}
p_{1,2}^j &= \frac{e^{\phi_{1,2}^j}}{e + e^{\phi_{1,2}^j}}, j = 1, 2, ..., N \\
\mathbf{p_{1,2}} &= [p_{1,2}^1, p_{1,2}^2, ..., p_{1,2}^N]^\mathbf{T}
\end{split}
\end{equation}

Next, subtract $\mathbf{p_{1,2}}$ from $\mathbf{a}$ to get
a vector $\mathbf{s_{1,2}} \in \mathbb{R}^{N \times 1}$
that represents differentiated scores between channels,
and upsampling $\mathbf{s_{1,2}}$ to the same size
as $\mathbf{W_2}$ to get the differentiated score matrix $\mathbf{S_{1,2}} \in \mathbb{R}^{N \times T}$, i.e.,
\begin{equation}
\begin{split}
\mathbf{s_{1,2}} &= \mathbf{a - p_{1,2}} \\
\mathbf{S_{1,2}} &= [\mathbf{s_{1,2}}, \mathbf{s_{1,2}}, ..., \mathbf{s_{1,2}}]
\label{eq:s12}
\end{split}
\end{equation}

Finally, the differential spatial information $\mathbf{W_{cd}}$
between channels can be extracted by multiplying
$\mathbf{W_2}$ by $\mathbf{S_{1,2}}$ as
\begin{equation}
\mathbf{W_{cd}} = \mathbf{W_2} \odot \mathbf{S_{1,2}}
\end{equation}
where $\mathbf{W_{cd}} \in \mathbb{R}^{N \times T}$
represents how much the differential spatial information
can be provided by the second channel over the first channel,
$\odot$ denotes
element-wise multiplication. Besides, to better guide the network
to extract the speech of the target speaker, we also
perform the target speaker adaptation on $\mathbf{W_{cd}}$
to exploit the target speaker-dependent spatial information, as
shown in Fig.\ref{fig:cdb_related} (b).

\section{EXPERIMENTS AND RESULTS}
\label{sec:exp_res}

\subsection{Dataset}
\label{ssec:datasets}

Our experiments are performed on the publicly available
multi-channel reverberant WSJ0-2mix corpus \cite{dataset}.
Multi-channel recordings are generated by convolving clean speech
signals with room impulse responses simulated with the image method
for reverberation time of up to about 600ms \cite{td_speakerbeam}.
The dataset consists of 8 channel recordings, but
to have a fair comparison with the state-of-the-art
baselines, we also use only two channels in our experiments.

We use the same way as in \cite{xu} to generate adaptation
utterances of the target speaker. The adaptation utterance is
selected randomly that different from the utterance in the mixture.
The adaptation recordings used in our experiments are anechoic.
The size of the training, validation, and test sets
are 20k, 5k, and 3k utterances, respectively.
All of the data are resampled to 8kHz from the original
16kHz sampling rate.

\subsection{Configurations}
\label{ssec:config}

Our experiments are performed based on the open source
Conv-TasNet implementation \cite{funcwj}. We use the same hyper-parameters as our baseline TD-SpeakerBeam in \cite{td_speakerbeam}.
%In the mask estimator, we use the channel-wise layer normalization (cLN)
%in the first convolution block and the global layer normalization (gLN) in other convolution blocks.

The same multi-task loss function in Equation (\ref{eq:multi}) of
TD-SpeakerBeam is used in our experiments. We set $\alpha=0.5$
to balance the loss tradeoff between the SiSDR and cross-entropy of SI-loss.
For those experiments with IPD combination, the IPD features are extracted
using an STFT window of 32 msec and hop size of 16 msec.
For the performance evaluation, both the
signal-to-distortion ratio (SDR) of BSSeval \cite{sdr} and
the SiSDR \cite{sisdr} are used. For more details of
experimental configurations, please refer to our released
source code on GitHub mentioned in the introduction.

\subsection{Results and Discussion}
\label{ssec:results}

\subsubsection{Baselines}
\label{baseline}

Two state-of-the-art multi-channel TSE systems are taken
as our baselines. One is the TD-SpeakerBeam with IPD \cite{td_speakerbeam}
and the other is the parallel encoder proposed in \cite{para}.
Results are shown in Table \ref{tab:base}.
System (1) and (2) are the results given in \cite{td_speakerbeam}.
As the source code of TD-SpeakerBeam and parallel encoder are not publicly available,
we implemented them by ourselves and reproduced the results of the system (3) to (6)
on the same WSJ0-2mix corpus. It's clear to see that
our reproduced results are slightly better than the results in
\cite{td_speakerbeam},
both the multi-task loss and
IPD are effective to improve the multi-channel TSE,
and it is interesting to find that only slight improvements are obtained
by using IPD together with the multi-task loss.

\begin{table}[h]
  \caption{SDR (dB) and SiSDR (dB) performances of our baselines.
  ``Multi" represents the multi-task loss. ``IPD" represents the
  system with internal combination of IPD features for
  multi-channel TSE. ``Parallel" represents the system with
  parallel encoder architecture.}
  \label{tab:base}
  \centering
  \begin{tabular}{l|c|c|c|c}
    \toprule
	System & Multi & IPD & SDR & SiSDR \\
    \midrule
	(1) TD-SpkBeam \cite{td_speakerbeam} & -          & - & 11.17 & - \\
	(2)    & -         & \checkmark  & 11.45 & - \\
    \midrule
	(3) TD-SpkBeam (our)  & -          & - & 11.26 & 10.76 \\
	(4)    				& \checkmark                  & -  & 11.51  & 11.00 \\
	(5)    				& \checkmark         & \checkmark  & 11.57  & 11.07  \\
	 \midrule
	(6) Parallel (our)    				& \checkmark         & \ -  & 12.43  & 11.91  \\
  	\bottomrule
  \end{tabular}
\end{table}

Furthermore, as we can see, the results of system (6)
are much better than the ones of other systems.
Actually, the original idea of the parallel encoder in \cite{para}
was proposed for the multi-channel speech separation, here
we extend it with TD-SpeakerBeam to improve the target speech extraction task.
It means that the parallel encoder is much effective than IPD features to capture the spatial information between
multi-channels. This may due to the parallel encoder is a
purely data-driven method and it is more suitable for the time-domain
TSE architecture.
We take the best results
from the system (5) and (6) in Table \ref{tab:base} as
our baselines.

\subsubsection{Results of the proposed methods}
\label{cd}

Table \ref{tab:cdcc} shows the performance comparisons
between TD-SpeakerBeam based TSE systems with different
multi-channel spatial information utilization techniques.
System (1) and (2) in this table are the best baselines.
It's clear to see that system (3) achieves both
2.4\% relative improvements in SDR and SiSDR over the system (2).
It indicates that performing a target speaker adaptation
on the multi-channel encoded representation can provide
more effective target speaker-dependent spatial information
than directly summing the multi-channel encoder outputs.

\begin{table}[ht]
  \caption{SDR (dB) and SiSDR (dB) performances of the proposed methods.
   ``Adapt" represents the system with the adaptation block in Fig.\ref{fig:cdb_related}.
    ``CD" represents the proposed Fig.\ref{fig:cdb_related} (b) with
    channel decorrelation. ``CC" represents the ``CD" block in Fig.\ref{fig:cdb_related} (b)
    is replaced by a channel correlation. Bold-fonts indicate best performance.
   All the systems are jointly trained with the multi-task loss.}
  \label{tab:cdcc}
  \centering
  \begin{tabular}{l|c|c|c|c}
    \toprule
	System  & IPD  & Adapt & SDR & SiSDR \\
    \midrule
	(1) TD-SpkBeam (our)          & \checkmark  & -  & 11.57      & 11.07 \\
	 \midrule
	(2) Parallel (our)         & -  & -  & 12.43      & 11.91 \\
	(3)             & -  & \checkmark    & 12.73      & 12.20 \\
	 \midrule	
	(4) CD         & -  & -  &  \textbf{12.87}      & 12.34 \\
	(5)            & -  & \checkmark  & \textbf{12.87}      & \textbf{12.35} \\
	(6)         & \checkmark  & \checkmark  & 12.55      & 12.01 \\
	\midrule	
	(7) CC           & -  & \checkmark  &  12.66      & 12.13 \\
  \bottomrule
  \end{tabular}
\end{table}

Moreover, by comparing results of the system (2) and (4), the inter-channel differential spatial information extracted by the proposed channel decorrelation is much better
than the cross-channel correlation spatial information captured
in the parallel encoder. Further  3.5\% SDR and 3.6\% SiSDR relative
gains have been obtained.
However, unlike the target speaker adaptation
in (3), an adaptation of the decorrelation spatial information
almost does not bring any performance improvements. Interestingly,
we find that incorporating the hand-crafted IPD spatial
features degrade the performances a little bit when the CD mechanism
is used.
This may due to the spatial information mismatch between
IPD and the inter-channel differential spatial information
extracted by CD. Because the IPD is computed in the frequency domain,
while the channel decorrelation is performed in the time-domain.

In addition, instead of the proposed CD for inter-channel differential
information, we also tried to exploit the
inter-channel correlation (CC) information, which
is achieved by replacing the $\mathbf{s_{1,2}}$ with $\mathbf{p_{1,2}}$
in Equation (\ref{eq:s12}) in Section \ref{ssec:cd}.
Results are shown in the system (7), which are much worse than the
results of the system (5). It indicates that the extracted inter-channel
differential spatial information is more effective than the
correlation information for the multi-channel end-to-end TSE systems.

Actually, all of the system (2) to (7) can be regarded as extensions
of the TD-SpeakerBeam, therefore, we can conclude that
the best result (system (5)) of our proposed method significantly
outperforms the multi-channel TD-SpeakerBeam baseline
by 11.2\% and 11.5\% relative improvements in SDR and SiSDR, respectively.

\subsubsection{Visualization}
\label{sec:vis}

\begin{figure}[h]
  \centering
  \includegraphics[scale=0.7]{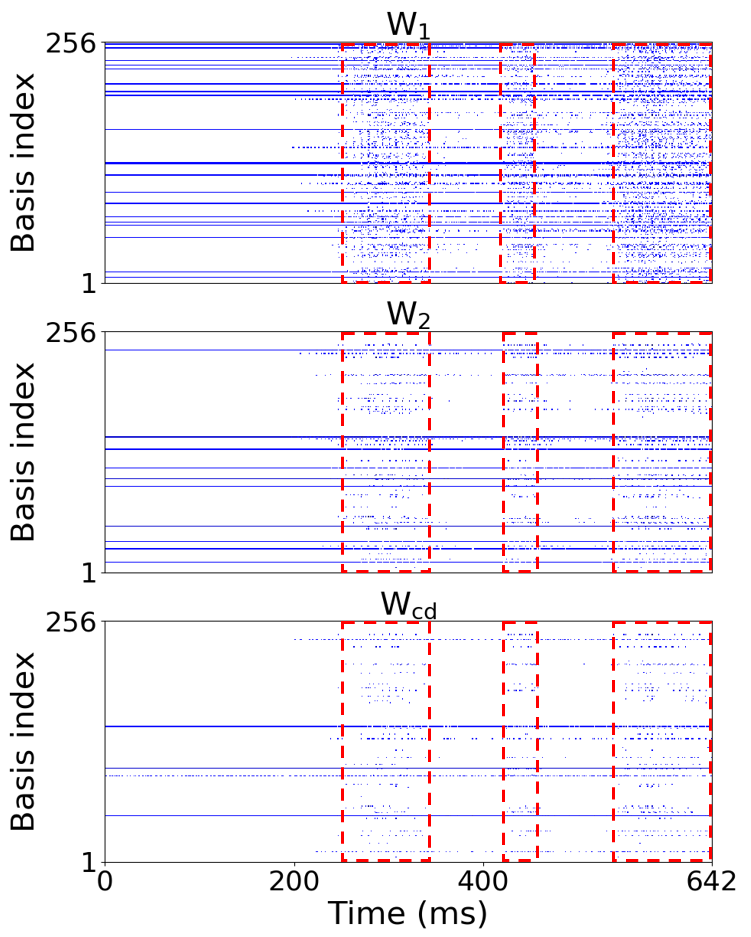}
  \caption{Visualization of the encoder and channel decorrelation representations of Fig.\ref{fig:cdb_related} (b).
The red dashed box indicates the most concentrated area.}
  \label{fig:visual}
\end{figure}

To better understand the role of each part during  the channel decorrelation,
we further investigate the distribution differences between the CD output
$\mathbf{W_{cd}}$ and its two encoded mixture
representation inputs $\mathbf{W_1}$ and $\mathbf{W_2}$ of one
mixture utterance with two overlapped speakers in Fig.\ref{fig:visual}.
As we expected, $\mathbf{W_1}$, $\mathbf{W_2}$, and $\mathbf{W_{cd}}$
have a similar pattern, and they all focus on the red dashed areas.
Given the nature of our task, we believe that the area is strongly related to the target speaker.
This means that after the whole TSE system training,
the network can automatically focus on the contents related to
the target speaker and ignore others.

Comparing the plot of $\mathbf{W_1}$ with $\mathbf{W_2}$,
their distributions are significantly different,
i.e., $\mathbf{W_1}$ is more densely distributed,
while $\mathbf{W_2}$ is more sparse. This indicates that
the $\mathbf{W_1}$ plays a main role during the target speech extraction, while
$\mathbf{W_2}$ plays a auxiliary role for providing
the complementary spatial information between channels.
Furthermore, by comparing the distribution of $\mathbf{W_2}$ with $\mathbf{W_{cd}}$,
some contents are removed.
Based on the calculation process described in Section~\ref{ssec:cd},
we believe that the removed information is the correlation
between $\mathbf{W_1}$ and $\mathbf{W_2}$,
only the inter-channel differential information is emphasized in $\mathbf{W_{cd}}$.

\section{CONCLUSION}
\label{sec:conclusion}

In this work, we propose two novel methods to extract the
multi-channel spatial information for the target speech extraction
task. Both methods are performed in the time-domain using
the end-to-end neural networks and they are extensions of the time-domain
SpeakerBeam.
One is designing a parallel encoder
with a target speaker adaptation layer to guide the target
speaker-dependent spatial information. The other is proposing
a channel decorrelation mechanism to effectively exploit the inter-channel
differential spatial information. Experiments on the reverberate WSJ0-2mix
corpus demonstrate that our proposed methods significantly
improved the multi-channel TD-SpeakerBeam for target speech extraction.
Our future work will focus on how to combine the hand-crafted
and data-driven based spatial features in an effective way.

\bibliographystyle{IEEEbib}
\bibliography{refs}

\end{document}